\documentclass{aa}
\usepackage{epsfig}
\usepackage{amsmath}
\usepackage{natbib}
\bibpunct{(}{)}{;}{a}{}{,}

\newcommand{\msun}{\mbox{$M_{\odot}$}}

\newcommand{\lsun}{\mbox{$L_{\odot}$}}

\newcommand{\mdot}{\mbox{$\dot{M}$}}

\newcommand{\msunyr}{\mbox{$M_{\odot} {\rm yr}^{-1}$}}

\newcommand{\kms}{km s$^{-1}$}

\def\lesssim{\mathrel{\hbox{\rlap{\hbox{\lower4pt\hbox{$\sim$}}}\hbox{$<$}}}}
\def\gtrsim{\mathrel{\hbox{\rlap{\hbox{\lower4pt\hbox{$\sim$}}}\hbox{$>$}}}}

\begin{document}

\title{Luminous Blue Variables as the progenitors of supernovae with quasi-periodic
       radio modulations}

\author{Rubina Kotak\inst{1} 
\and Jorick S. Vink\inst{2}}

\institute{European Southern Observatory, Karl-Schwarzschild Str. 2,
           Garching bei M\"{u}nchen, D-85748, Germany
\and Astrophysics group, Lennard-Jones Laboratories, Keele University, 
Staffordshire, ST5 5BG, UK 
           }
\offprints{R. Kotak, rkotak@eso.org}

\titlerunning{LBVs as progenitors of SNe with pronounced radio modulations}
\authorrunning{}

\abstract{The interaction between supernova ejecta and circumstellar
matter, arising from previous episodes of mass loss, provides us with 
a means with which to constrain the progenitors of 
supernovae.
Radio observations of a number of supernovae show quasi-periodic 
deviations from a strict power-law decline at late times. Although several 
possibilities have been put forward to explain these modulations, no single 
explanation has proven to be entirely satisfactory. Here we suggest that Luminous 
Blue Variables undergoing S-Doradus type variations give rise to enhanced phases 
of mass loss which are imprinted on the immediate environment of the exploding
star as a series of density enhancements. The variations in mass loss arise
from changes in the ionization balance of Fe, the dominant ion that drives
the wind. With this idea, we find that both the recurrence timescale 
of the variability, as well as the amplitude of the modulations are in line 
with the observations. Our scenario thus provides a natural, single-star 
explanation for the observed behaviour that is, in fact, expected on theoretical 
grounds.
}
\maketitle

%*****************************************************************

\section{Introduction}
\label{s_intro}

Massive stars and core-collapse supernovae (SNe) play an important role 
in driving the chemical evolution of galaxies and in shaping the dynamics 
of the interstellar medium over all cosmological epochs, since 
the very first stars came into existence. 
Despite their importance, the lives and deaths of massive 
stars are poorly understood; in particular, it is not known with any
degree of certainty which massive stars produce which SNe. 

While progress is gradually being made in the direct identification of 
core-collapse SN progenitors by searching for these in pre-explosion images 
\citep[e.g.][]{smartt:02,vdyk:03}, such studies are limited to either extremely 
nearby SNe, and/or those occurring in uncrowded regions.
Perhaps as a result of the stellar initial mass function, most direct
detections of progenitors have yielded masses not significantly greater than
$\sim$10$\msun$.  

Although the evolution of the most massive stars (with $M$$>$30\,\msun) is 
largely unconstrained, it is generally accepted that mass loss drives 
these objects through the O star, Luminous Blue Variable (LBV), and Wolf-Rayet (WR)
phases \citep[e.g.][]{chiosimaeder:86}.
Significant progress has been made in recent years in our 
understanding of the mass loss properties over the various evolutionary phases
of massive stars. In a series of papers, \citet{vink:99,vink:00,vink:01} provided 
predictions of mass-loss rates of O and B supergiants, LBVs \citep{vink:02}, and 
late-type WR-type stars as a function of metal content \citep{vink:05}.

SNe that result from the core-collapse of massive stars
explode in environments that have been modified by 
%UV radiation and 
mass loss from the progenitor. The SN ejecta interact first
with this circumstellar material before interacting with
interstellar material. We might therefore expect that the distinct 
differences in wind properties over the lifespan of a massive star might
be imprinted onto the resulting circumstellar media (CSM). 
Furthermore, we would then expect these differences to be manifested
in the interaction between the SN ejecta and surrounding material. By 
quantifying these differences one can then constrain the evolutionary 
phase of the exploding object. 

Over the last decades, radio observations of SNe have provided a means with 
which to constrain the density of the CSM around core-collapse SNe. 
%\citep[e.g.][]{weiler:86,weiler:02}. 
The inferred mass-loss rates
from modelling of most radio SN light curves yield values of 
$\mdot$ $\sim$$10^{-6}$ -- $10^{-4}$$\msunyr$ 
\citep[e.g. compilation in][]{weiler:02}. Unfortunately, these average mass-loss rates
are generally only accurate to within a factor of $\sim$10, and are typical of almost all 
types of massive star, making it difficult to pin down the evolutionary phase during
which core-collapse occurred.

A small subset of radio SNe, however, show quasi-periodic modulations in their radio 
lightcurves. In what follows, we argue that this type of modulation may be the result 
of an LBV star that underwent S~Dor variations which entailed opacity changes in the 
wind-driving region, resulting in varying mass-loss rates.

\section{LBV mass-loss variability}
\label{s_lbv}

LBVs are unstable massive stars, located in the upper portion ($L/\lsun \ga 5.5$) of 
the Hertzsprung-Russell diagram \citep{hd:94}. 
They exhibit different types of variability which can be divided into 
three categories: 
(i) small-amplitude ($\sim$0.1 mag) ``micro'' variability, which is common amongst blue supergiants, 
(ii)  ``moderate'' S~Doradus variations of $\sim$1 -- 2 mag (SD phases), and 
(iii) truly ``giant'' eruptions, 
of which the Galactic cases P~Cyg and Eta~Car are most well-known. 
%{\bf \citep[see][]{smith:06}. 
According to an extensive list compiled by \citet{vgenderen:01}, 
the SD-phases occur on two timescales: $<$\,10 yrs. (``short SD'' phases), 
and $>$\,20 yrs. (``long SD'' phases).

Spectroscopically-determined LBV mass-loss rates are found to change on 
S~Dor time-scales \citep[e.g.][]{stahl:01, stahl:03}.
\citet{vink:02} attributed this mass-loss variability to the ionization 
and recombination of the Fe lines which are responsible for driving the 
wind. This leads to ``bi-stable'' behaviour in the wind \citep{pp:90,lamers:95,vink:99}
causing the star to flip back and forth between two states: that of low mass 
loss, high-velocity winds, to high mass-loss, low velocity, winds. This bi-stable
behaviour results in jumps in the mass-loss rate of a factor of $\sim$2 for LBVs \citep {vink:02},
and a decrease by a factor of $\sim$2 in the terminal wind-velocity between O and B supergiants
\citep{lamers:95}. 
The wind density would therefore be expected to change by a factor of $\sim$4 on the 
timescale of the S Doradus variations. In the absence of any other material around 
the star, this would result in a pattern of concentric shells of varying density.

\section{Radio SN Lightcurves and stellar wind interaction} 
\label{s_rsne}

The properties of Radio SNe (RSNe) lightcurves and the model for the interaction 
of the SN blastwave with the surrounding CSM as described by \citet{chevalier:82a,chevalier:82b} 
has been reviewed by \citet{weiler:02}. In short, the source switches on, reaches a peak, 
and generally decays smoothly over a timescale of years. The peak is attained at different 
times (days to months) for different radio frequencies, depending on the optical depth of  
either external and/or internal absorbing material. For a discussion on the relative
importance of free-free absorption (FFA) by an external -- possibly clumpy -- medium, versus 
synchrotron self-absorption (SSA), we refer the reader to \citet{chevalier:98} and 
\citet{fb:98}. 
The radio emission nevertheless places constraints 
on the density $\rho$, and therefore on the ratio of \mdot\ to the terminal wind velocity 
($v_{\rm w}$) of the CSM of the SN progenitor: $\rho \propto \mdot/v_{\rm w}r^2$.
Using RSNe lightcurves, and generally assuming a typical $v_{\rm w}$ for a red 
(super)giant ($\sim$10--20\,\kms) constraints on \mdot\ of pre-SN winds have 
been placed on about a 
dozen RSNe \citep[see Table~3,][]{weiler:02} with typical \mdot\ values in the range 
$10^{-6}$--$10^{-4}$ \msunyr. 

The shock radius can be constrained through radio observations, whilst 
the measured variability timescale is set by the wind velocity 
of the progenitor (see below). Ideally, one would wish to spatially resolve the 
shock radius directly, as was possible for SN~1979C using VLBI techniques \citep{bb:03}. 
Usually, one needs to assess the relative importance of SSA vs. FFA.
If SSA prevails, and if the peak flux is available, i.e., at the transition between
the optically thick to thin phases, assuming equipartition between the magnetic 
fields and relativistic electrons yields an estimate of the size of the radio-emitting
region \citep{shklovskii:85,chevalier:98}. 
While if FFA is dominant, it is usual to simply adopt the measured velocity from 
optical lines at early times together with parameters from modelling the light curves
at different frequencies \citep{weiler:02}.

\section{Modulations in radio SNe lightcurves}

Variations in the radio lightcurves have been reported for about a dozen SNe 
of both type II and type Ibc (see Table 4 of Soderberg et al. 2005). These can generally 
be interpreted as due to variations in the mass-loss rate due to a changing evolutionary 
phase of the SN progenitor. A subset of radio SNe show episodic bumps.
These include the classic cases of SN~1979C \citep{weiler:92}
and SN~1998bw \citep{kulkarni:98} that was associated with GRB~980425 \citep{galama:98}, 
as well as two recent radio SNe, \object{SN~2001ig} \citep{ryder:04} 
and \object{SN~2003bg} \citep{soderberg:05}. This has brought the number of radio 
SNe that show recurrent modulations in their radio light curves, to four.
A caveat to the above is that finer sampling of radio light curves may reveal 
the extent to which such behaviour is commonplace.
Interestingly, SNe 2001ig and 2003bg are strikingly similar both in terms of the 
amplitudes and timescales of the variations 
(see Fig. 8 in Soderberg et al. 2005 for a comparison). We focus our discussion
on these two SNe, and defer a discussion of \object{SN~1979C}
and 1998bw to \S \ref{sec:other}. 

\subsection{SN~2001ig}
\label{sec:01ig}

SN~2001ig is a transition object: on the basis of optical spectroscopy, 
it was initially classified as type II (i.e. showing H lines) by
\citet{matheson:01} but metamorphosed into type Ib/c object (i.e. no H 
lines, weak He lines) by \citet{filippenko:02} about 9 months later.
This suggests that it has lost most of its H-rich envelope \citep{ryder:04}.
We will return to this point in \S \ref{s_disc}.

For SN~2001ig, \citet{ryder:04} used the model of \citet{weiler:02} to
derive parameter fits to the radio lightcurves yielding parameters such 
as the radio spectral index $\alpha$, the decline rate of the optically 
thin phase $\beta$, and importantly, at least one parameter $\delta$ (= $-2.56$) 
that describes how the optical depth of the local CSM changes with time. 
The relevance of $\delta$ is that it relates to the rate at which the SN blast-wave 
radius $r_{\rm BW}$ changes with time:
$r_{\rm BW} \propto {t}^{m}$ with $m = -\delta/3.$
In the mini-shell model by \citet{chevalier:82a,chevalier:82b}, 
$m=1$ implies a non-decelerating CSM. For SN~2001ig, $m$ = 0.85. 

The recurrence timescale ($t$) of the bumps in the radio flux of 
SN~2001ig is derived to be 150 days. Using Eq.~(13) from \citet{weiler:86}:

\begin{equation}
\Delta P~=~\frac{R_{\rm shell}}{v_{\rm w}}~=~\frac{v_{\rm ejecta}~t_{\rm i}}{v_{\rm w}~m} \left(\frac{t}{t_{\rm i}}\right)^m
\label{eq:period}
\end{equation}
where $t_{\rm i}$ is set by when the ejecta velocity was measured. 
For SN~2001ig, $t_{\rm i}$\,$\sim$\,14\,d and $v_{\rm ejecta}$ $\sim$15000\,km\,s$^{-1}$
\citep{cp:01}.  We stress that the dependence on $t_i$ is weak. 
Assuming $v_{\rm w}$ = 10--20\,km\,s$^{-1}$, typical wind velocities for 
red (super)giants, \citet{ryder:04} found a period $P$ between successive 
mass-loss episodes that was too long for red (super)giant pulsations (typically 
a few 100 days), and too short for thermal pulses (10$^2$--10$^3$ years). 
For these reasons, they invoked an edge-on, eccentric binary scenario,
consisting of a WR-star and a massive companion. The configuration of the binary
system would lead to a pile-up of mass, giving rise to modulations in the 
radio light curve. Such a scenario was proposed on the basis of 
hydrodynamical simulations by \citet{sp:96} to explain the radio lightcurve
of SN~1979C. 

One of the main differences between LBV and red giant winds is that 
LBV winds are faster, typically 100--500\,km\,s$^{-1}$ \citep[e.g.][]{leitherer:97}
c.f.  10--20\,\kms\ for red (super)giants. This implies that if the progenitor of 
SN~2001ig were an LBV, the expected period between successive 
mass-loss episodes would be $\Delta P \sim$\,25\,yr 
(for an assumed $v_{\rm w}$=200\,\kms), consistent with the long SD phase.
Inconsistencies in the derivation of $\Delta P$
in \citet{ryder:04} results in their inferred radial spacings between 
the shells being too small ($\sim 2\times10^{15}$\,cm, with $t_i=1$\,d c.f. $1.6\times10^{16}$\,cm 
from above) and might therefore be incompatible with the kind of 
WR pinwheel nebula they describe.

\subsection{SN~2003bg}

SN~2003bg was classified as a type Ic (no H, He, Si lines) by \citet{filippenko:03}; 
in less than a month, the spectrum of SN~2003bg had evolved to a type II SN, showing
broad P Cygni lines of hydrogen \citep{hamuy:03}. \citet{soderberg:05} are able to 
fit the the radio light curve under the assumption of SSA.

Remarkably, the short-term variability in the multi-frequency radio lightcurves
of SN~2003bg, is found to be on very similar timescales as SN~2001ig. \citet{soderberg:05}
infer enhancements of a factor of $\sim\negthickspace2$ in density during the deviations 
from pure power-law evolution.
They consider a range of possibilities that might account for the observed modulations 
in the lightcurve, but favour a single-star progenitor model of a WR star that underwent 
episodes of intensified mass loss. However, they do not specify the nature of the physical 
mechanism that might give rise to such periods of enhanced mass loss.
Using the measured expansion velocity for SN~2003bg at a time $t_{\rm i}$ $\sim$\,20 
days \citep{filippenko:03,hamuy:03}, we find $\Delta P$ $\sim$\,25 years, 
again, consistent with the long SD phase.

%Using equipartition analysis, they infer spatial scales of the density 
%enhancements of approximately $4 \times 10^{16}$ and $8 \times 10^{16}$cm, 
%which are larger than would be inferred assuming FFA only \citep{ryder:04}.
%However, even the larger spatial scales deduced from SSA analysis would yield
%values of $\Delta P$ encompassed by current LBV variability studies, although
%these would be at the upper end, i.e., decades.

\subsection{The cases of SN~1979C and SN~1998bw}
\label{sec:other}

As outlined above, the relevance of SNe 2001ig and 2003bg-like events is that 
there is evidence for regularly changing mass loss of the SN progenitor. 
These observations may finally open up the possibility of yielding more stringent 
constraints on the pre-SN mass-loss properties than available from 
previous radio data.
To our knowledge, the only massive stars known to fulfil the
criterion of quasi-regularly spaced mass-loss episodes are LBVs.

The only other SNe that have shown episodic modulations in their
radio lightcurves are SN~1979C and SN~1998bw. 
For SN~1979C, \citet{weiler:92} found a radio modulation on a timescale of 
$\sim$1575 days, i.e. a factor $\sim$10 longer than for SNe 2001ig/2003bg. However, 
when we account for the rate at which the blast-wave radius changed with 
time, $\delta = -2.94$ \citep{montes:00}, we find $\Delta P\sim$\,160\,yrs for
$v_{\rm{ejecta}}$ $\sim$\,8000\,\kms \citep{panagia:80}.
Now the circumburst radius of SN~1979C has been resolved with VLBI.
Using the value thus derived of the radio size \citep[Fig. 5,][]{bb:03}, 
combined with the epoch of the radio modulations at $\sim$4 and 8 years, 
we find a difference in angular size of $\sim$0.5\,mas. 
Assuming a distance of 15.2\,Mpc \citep{freedman:01}, then implies 
mass loss variability on timescales of $\sim$150 years ($v_{\rm w} = 200$\,\kms), 
i.e.  consistent with $\Delta P$ derived above, but longer than for SNe 2001ig/2003bg. 
We note as an aside that \citet{vgenderen:01} does list some LBVs with SD phases that 
are greater than $\sim$100 years.

Interestingly, for the more exotic case of SN~1998bw, a similar analysis as 
in \S\,\ref{sec:01ig}, with $m$ = 0.78 \citep{weiler:01}, and a timescale of 
$t\simeq50$ days, we find $\Delta P$\,$\sim$\,6\,years, which 
is consistent with the short SD timescale. We note however that the episodic 
bumps in the lightcurve of SN~1998bw are not as well-defined as the sinusoidal 
variations of SNe 2001ig/2003bg. 

\section{Discussion}
\label{s_disc}

We have proposed that LBVs undergoing S Doradus-type variations are viable progenitors 
of SNe that show quasi-sinusoidal variations in their radio lightcurves. 
The main motivation for this suggestion is that S Doradus-type winds provide a natural 
and physical mechanism that could give rise to the observed radio modulations. 
Both the timescales and the amplitude of the variations in CSM density as a result 
of the S Doradus variations are in line with the constraints imposed by the
radio SN data. Our proposed scenario therefore provides support for a 
single-star progenitor system.
Admittedly, there are uncertainties associated with current analyses of the 
timescales of the radio SNe described in this paper, as well as with LBV S~Dor 
variations in general. Until a significant body of radio SNe with episodic 
modulations becomes available, statistical considerations are deemed premature. 
If shorter SD variations are superimposed on the longer ones, then 
these might be revealed by further increasing the cadence of radio
observations to a nightly basis.

It may be relevant that both SNe 2001ig and 2003bg are transitional objects. 
SN~2001ig is an example of the group of IIb SNe, that switch from type II 
(H and He both present) to Type Ib (He present, H absent), whilst 2003bg was first
classified as a type Ic, but then underwent a change to a type II. 
These facts suggest that SN2001ig and 2003bg have a rather limited amount of H. 
This is also consistent with an LBV scenario, as LBV atmospheres are H-rich 
compared to WR stars, but He-rich compared to OB stars and red supergiants.

We do not exclude the possibility that objects like SNe 2003bg and 1998bw are 
in fact ``stripped core'' WR stars by the time they explode. However, given that 
WR stars are not known to exhibit the type of variability that we consider here,
we believe it is more likely that the observed quasi-sinusoidal variations in their 
radio lightcurves are attributable to their prior LBV phase. This would mean that 
these objects are WR stars for only a very brief amount of time before they explode, 
and to all intents and purposes, we can categorize these objects as LBVs.

The fact that our proposed scenario is simple and workable does not mean other 
scenarios need necessarily be excluded. \citet{ryder:06} recently found 
evidence for a binary companion of SN~2001ig. This in itself is not surprising; 
it is well-known that a large fraction of massive stars occur in binary systems 
\citep[e.g.][]{mason:98}.  The more relevant question of whether the potential
binary companion ever had any direct influence on the star that exploded as 
SN~2001ig is more difficult to answer. 

The binary scenario that could give rise to a pinwheel nebula as proposed by 
\citet{ryder:04,ryder:06} and \citet{sp:96} requires the binary to be eccentric
and viewed close to edge-on. 
Given the striking similarity of SN~2003bg to SN~2001ig, 
\citet{soderberg:05} argue that the Wolf-Rayet pinwheel nebula hypothesis is unlikely, 
as the configurations of both systems must necessarily be very similar. They therefore 
propose a single WR progenitor, but do not put forward a mechanism for the CSM variability. 
The LBV progenitor scenario presented here may remove this shortcoming.

\begin{acknowledgements}
We thank the anonymous referee for constructive criticism, and Dr. S. Ryder 
for an interesting seminar on SN~2001ig which stimulated this work.
JSV acknowledges financial support from an RCUK fellowship.
RK acknowledges support from an ESO fellowship.
\end{acknowledgements}

\end{document}